\documentclass[conference]{IEEEtran}

\usepackage{amsmath}
\usepackage{cases}
\usepackage{graphicx}

\usepackage{cite}
\usepackage{times}

\usepackage{graphicx}
\usepackage{graphics}
\usepackage{subfigure}
\usepackage{epsfig}
\usepackage{latexsym}
\usepackage{amsfonts}
\usepackage{amssymb}
\usepackage{paralist}
\usepackage{xspace}
\usepackage{mathrsfs}
\usepackage{amssymb}
\usepackage{slashbox}
\usepackage{color}
\usepackage[ruled]{algorithm}
\usepackage[noend]{algorithmic}

\usepackage{amssymb}
\usepackage{url,epsfig,array}

\allowdisplaybreaks 
\def\ie{\textit{i.e.}\xspace}

\def\eg{\textit{e.g.}\xspace}

\newcommand{\CUTXY}[1]{{}}

\renewcommand{\paragraph}[1]{\smallskip \noindent {\textbf{#1}}}
\newcommand{\tabincell}[2]{\begin{tabular}{@{}#1@{}}#2\end{tabular}}

\newtheorem{theorem}{Theorem}

\newtheorem{definition}{Definition}

\begin{document}

\title{Message in a Sealed Bottle: \\Privacy Preserving Friending in Social Networks}

\author{\authorblockN
{Lan Zhang\authorrefmark{1},
Xiang-Yang Li\authorrefmark{2}}
\authorblockA{\authorrefmark{1} Department of Software Engineering,
  Tsinghua University}
\authorblockA{\authorrefmark{2} Department of Computer Science,
  Illinois Institute of Technology, and TNLIST, Tsinghua University}}
\maketitle

\begin{abstract}
Many proximity-based mobile social networks are developed to facilitate connections between any two people, or to help a user to find people with matched profile within a certain distance. A challenging task in these applications is to protect the privacy  the participants' profiles and personal interests.

In this paper, we design  novel mechanisms, when given a preference-profile submitted by a user, that search a person with matching-profile  in decentralized multi-hop mobile social networks. Our mechanisms are privacy-preserving:  no participants' profile and the submitted preference-profile are exposed.
Our mechanisms establish a secure communication channel between the initiator
 and matching users at the time when the matching user is found.
Our rigorous analysis shows that our mechanism is secure, privacy-preserving, verifiable, and efficient both in communication and computation.
Extensive evaluations using real social network data, and actual system implementation on smart phones show that  our mechanisms are significantly more efficient then existing solutions.
\end{abstract}


\section{Introduction}
\label{sec:introduction}

A boom in mobile hand-held devices greatly enriches the social networking applications.
Many social networking services are available on mobile phones (\eg, JuiceCaster and
MocoSpace) and majority of them are location-aware (\eg, FourSquare, BrightKite and Loopt).
However, most of them are designed for facilitating people connection based
on their real life social relationship.
There is an increasing difficulty of befriending new people
or communicating with strangers while protecting the privacy of real personal information.

Friending and communication are two important basic functions of social networks.
When people join social networks,
 they usually begin by creating a profile, then interact with other users.
The content of profile could be very broad,
 such as personal background, hobbies, contacts, places they have been to.
Profile matching is a common and helpful way to make new friends with common interest or experience,
 find lost connections, or search for experts.
Some applications  help a user automatically find  users
 with similar profile within a certain distance.
For example, in the proximity-based mobile social network "\emph{Color}",
 using the GPS and Bluetooth capabilities on phones,
 people in close proximity (within 50 meters) can share photo automatically based on their similarity.
MagnetU\cite{MagnetU}
 matches one with nearby people for dating, friend-making.
Small-talks\cite{yang2010smalltalker} connects proximate users based on common interests.
These applications use profiles to facilitate friending between strangers and
 also enable privacy preserving people searching to some extent.
Observe that in practice the mobile Internet connection may not always be available
 and it may incur high expense.
Thus, the availability of short-range wireless technology such as WiFi
 and Bluetooth makes it possible to build a new class of proximity-based
 decentralized (or ad hoc) social networks with mobile
 phones \cite{Patrick2009enabling,zhang2010wiface}.

However the increasing privacy concern becomes a barrier for mobile ad hoc social networking.
People are unwilling to disclose personal profiles
to arbitrary persons in physical proximity before deciding to interact with them.
The insecure wireless communication channel
and potentially malicious service provider increase
the risk of revealing private information.
Friending based on \emph{private profile matching} allows two users to
 match their personal profiles without disclosing them to each other.
There are two mainstreams of approaches to solve the privacy-preserving profile-based friending problem.
The first category treats a user's profile as a set of attributes
 and provide well-designed protocols to privately match users' profiles based on
 \emph{private set intersection} (PSI) and \emph{private cardinality of set intersection}  (PCSI),
 \cite{von2008veneta,li2011findu}.
The second category considers a user's profile as a vector and measures the
 social proximity by \emph{private vector dot product} \cite{ioannidis2002secure,dong2011secure,zhang2012fine}.
They rely on public-key cryptosystem and homomorphic encryption,
 which results in expensive computation cost and usually requires a trusted third party.
Multiple rounds of interactions are required to perform the public key exchange
 and private matching between each pair of parties, which causes high communication cost
 in mobile social networks.
Presetting (\eg exchange public keys) is often required by these approaches before matching.
In the final step of these protocols, only one party learns the matching result, which makes them \emph{unverifiable}.
And there lack efficient methods to verify the result.
Moreover, in these approaches, matched users and unmatched users
 all get involved in the expensive computation and learn the matching result (\eg profile intersection)
 with the initiator despite  different similarities between them.
Most of them are vulnerable to active attacks like dishonesty and colluding adversary.
These limitations hinder the adoption of the SMC-related private matching methods in mobile social networks.

Furthermore, a secure communication channel is also equally important in mobile social networks.
Although the matching process is private, the following chatting
may still be disclosed to adversary and more privacy may be leaked.
Most work assume that there is a secure communication channel established
 by using public key encryption system.
This involves a trusted third party and key management,
  which is difficult to manage in decentralized mobile social networks.

Facing these challenges,
 we first formally define the privacy preserving verifiable computation problem (Section \ref{sec:prodef}).
We then propose several protocols (Section \ref{sec:secretshare}) to address the privacy preserving
 profile matching and secure communication channel
 establishment in decentralized social networks without
 any presetting or trusted third party.
We take advantage of the common attributes between matching users, and use it
 to encrypt a  message with a secret key in it.
In our mechanisms, only a matching user can decrypt the message, and unmatched users learn nothing.
A privacy-preserving profile
 matching and secure channel construction are completed simultaneously with one round communication.
The secure channel construction resists the man in the middle attack.
A sequence of well-designed schemes make our protocol practical, flexible and lightweight, \eg,
  a remainder vector is designed to significantly reduce the computation and communication cost of unmatched users.
Our profile matching mechanism is also \emph{verifiable} which thwarts cheating about matching result.
Both precise and fuzzy matching/search in a flexible form are supported.
The initiator can define a similarity threshold, the participants whose similarity is below the threshold
 learns nothing.
We also design mechanism for location privacy preserved vicinity search based on our basic scheme.
Compared to most existing work (Section \ref{sec:review}) relying on the asymmetric cryptosystem and trust third party,
 our protocols require no presetting and much less computation.
To the best of our knowledge, these are the first privacy-preserving verifiable profile matching protocols
  based on symmetric cryptosystem.

We rigorously analyze the security and performance of our mechanism (Section \ref{sec:analysis}).
We then conduct extensive evaluations on the performances of our mechanisms using large scale social network data, Tecent Weibo. Our evaluation results (Section \ref{sec:evaluation}) show that our mechanisms outperform existing solutions significantly.
We also implement our protocols in laptop and mobile phones and measure the computation and communication cost in real systems.
In our mobile-phone implementation, a user only needs about 1.3ms to generate a friending request.
For non-candidate users, on average, it takes about 0.63ms to process this friending request, while for candidate users, on average it takes about 7ms to process this request.

\section{System Model and Problem Definition}
\label{sec:prodef}
\subsection{System Model}
\label{sec:model}

We consider a mobile ad hoc social networking system.
When a person joins a social network,
 he/she usually begins by creating his/her own profile
 with a set of attributes.
The attribute can be anything generated by system or input by users,
 including his/her location, places he/she has been to,
 his social groups, experience, interests, contacts, keywords of his/her blogs, etc.
According to our analysis of two well-known social networking systems (Facebook and Tecent Weibo),
more than $90\%$ users have unique profiles.
Thus for most users, the complete profile can be his/her fingerprint in social networks.
Then, in most social networks, friending usually takes two typical steps:
 profile matching and communication.
These  applications cause a number of privacy concerns.
\begin{enumerate}
\item \textbf{Profile Privacy}: The profiles of all the participants,
    including the initiator, intermediate relay users and the matching target,
    should not be exposed without their consent.
    For the initiator, the required profile
    could be his/her own profile or his/her desired person' profile.
    For other participants, \eg, the unmatched relay users and the matching users,
    protecting their profiles is necessary and can reduce the barrier
    to participate in the mobile social networks (MSN).
    Note that, the exact location information
    is also a part of the user's profile privacy.

\item \textbf{Communication Security}: The messages between a pair of users should be
    transmitted through a secure communication channel.
    We emphasize that the secure communication channel establishment has
    been ignored in most previous work which address the private profile matching in
    decentralized mobile social networks.
    In practice, after profile matching, more privacy, even profile information,
    may be exposed to adversaries via communication through an insecure channel.
\end{enumerate}

%

In this paper, we address the privacy preserving
 profile matching and secure communication channel
 establishment in decentralized social networks without
 any presetting or trusted third party.
We define the problem formally before presenting our mechanism.
Each user $v_k$ in a social network has a profile which is a set
 consisting of $m_k$ attributes,
 $A_k= \{a_k^1, a_k^2, ... , a_k^{m_k}\}$.
The number of attributes is not necessary
 the same for different users.
Each attribute consists of a header indicating its category name
 and a value field with a single value or multiple values.
An initiator $v_i$ represents his/her desired user by
 a request attribute set with $m_t$ attributes as
 $A_t= \{a_t^1, a_t^2, ... , a_t^{m_t}\}$.
Our mechanism allows the initiator to search a matching user
 in a flexible way by constructing the request attribute set
 in the form of $A_t=(N_t, O_t)$. Here
\begin{itemize}
\item $N_t$ consists of $\alpha$ \emph{necessary} attributes.
 All of them are required to be owned by a matching user;
\item $O_t$ consists of the rest $m_t-\alpha$ \emph{optional} attributes.
 At least $\beta$ of them should be owned by a matching user.
\end{itemize}
The acceptable \emph{similarity threshold} of a matching user is
$\theta  = \frac{\alpha + \beta}{m_t}$.
Let $\gamma = m_t-\alpha-\beta$.
When $\gamma = 0$, an perfect match is required.
A \emph{matching user} $v_m$ with a profile $A_m$ satisfies that
\begin{equation}
N_t \subset A_m \text{ and } |O_t \cap A_m| > \beta.
\end{equation}
When $A_t \subset A_m$, $v_m$ is a perfect matching user.
In a decentralized mobile social network, a request
 will be spread by relays until hitting a matching user
 or meeting a stop condition, \eg expiration time.
When a matching user is found, the initiator $v_i$ and the match user $v_m$
 decide whether to connect each other.

\subsection{Adversary Model}
\label{sec:adversary}

In the profile matching phase,
 if a party obtains one or more users¡¯
 (partial or full) attribute sets without the explicit consents from those users,
 it is said to conduct \emph{user profiling} \cite{li2011findu}.
Two types of user profiling are taken into consideration.
In the \emph{honest-but-curious} (HBC) model, a user tries to learn more
 profile information than allowed by inferring from the information
 he/she receives and relays but honestly follow the mechanism.
In a \emph{malicious} model, an attacker deliberately deviates from the mechanism to
 learn more profile information.
In this work we  consider several powerful malicious attacks.

\begin{definition}[Dictionary profiling]
 A powerful attacker who has obtained the dictionary of all possible attributes
  tries to determine a specific user's attribute set by enumerating or guessing all likely
  attribute sets.
\end{definition}
Most existing private profile matching approaches
 are vulnerable to the dictionary profiling attack.

\begin{definition}[Cheating]
Cheating is another threat for most existing private profile matching approaches.
In the process of profile matching, a participant may cheat by deviating from the agreed protocol.
\end{definition}

In the communication phase, an adversary can learn the content of messages
 by eavesdropping. The construction of a secure channel may suffer the Man-in-the-Middle (MITM) attack.
There are other saboteur attacks.
In the deny of service (DoS) attack, an adversary keeps sending profile matching requests.
The DoS attack can be prevented by restricting the frequency of relay and reply requests
 from the same user.
Note that, some saboteur behaviors are precluded, such as alter or drop the requests or replies.

\subsection{Design Goal}


The main goal and great challenge of our mechanism is
 to conduct efficient matching against the user profiling and cheating,
 as well as establish a secure communication channel thwarting
 the MITM attack and eavesdropping in a decentralized manner without a trust third party.
In our mechanism, a user can define a similarity threshold to
 protect his/her privacy from the user whose similarity is not up to the threshold.
Specifically, We define different privacy protection levels $PPL(A_k,v_j)$ of
 a profile $A_k$ of $v_k$ against a user $v_j$.
\begin{definition}[Privacy Protection Level] Four different privacy protection levels
(PPL0, PPL1, PPL2, PPL3) will be discussed  in this work. Specifically,

\textbf{PPL0}: If $PPL(A_k, v_j)=0$, $v_j$ can learn the  profile $A_k$.

\textbf{PPL1}: If $PPL(A_k, v_j)=1$, $v_j$ can learn the intersection of $A_k$ and $A_j$.

\textbf{PPL2}: If $PPL(A_k, v_j)=2$, $v_j$ can learn the  $\alpha$ necessary attributes of $A_k$ and
 the fact that at least $\beta$ optional attributes are satisfied.
 Specially, when $\alpha=0$, $v_j$ learns the fact
  that the cardinality of $A_k \bigcap A_j$ exceeds the threshold.

\textbf{PPL3}: If $PPL(A_k, v_j)=3$, $v_j$ learns nothing about $A_k$.
\end{definition}

\medskip

We design our mechanism to achieve PPL2 against matching user and PPL3 against unmatching user
 in both HBC and malicious model and thwart cheating.
We also optimize the mechanism to reduce the computation cost for unmatched users.
Furthermore, in our mechanism human interactions are needed
 only to decide whether to connect their matching users.

\section{Privacy Preserving Profile Matching and Secure Communication}
\label{sec:secretshare}

Here we present our lightweight privacy preserving flexible profile matching
 in decentralized mobile social networks without any presetting or trust third party.
A secure communication channel is constructed
 between matching users.

\begin{figure*}[hptb]
\begin{minipage}[b]{0.45\linewidth}
\includegraphics[width=0.95\textwidth,height=1.7in]{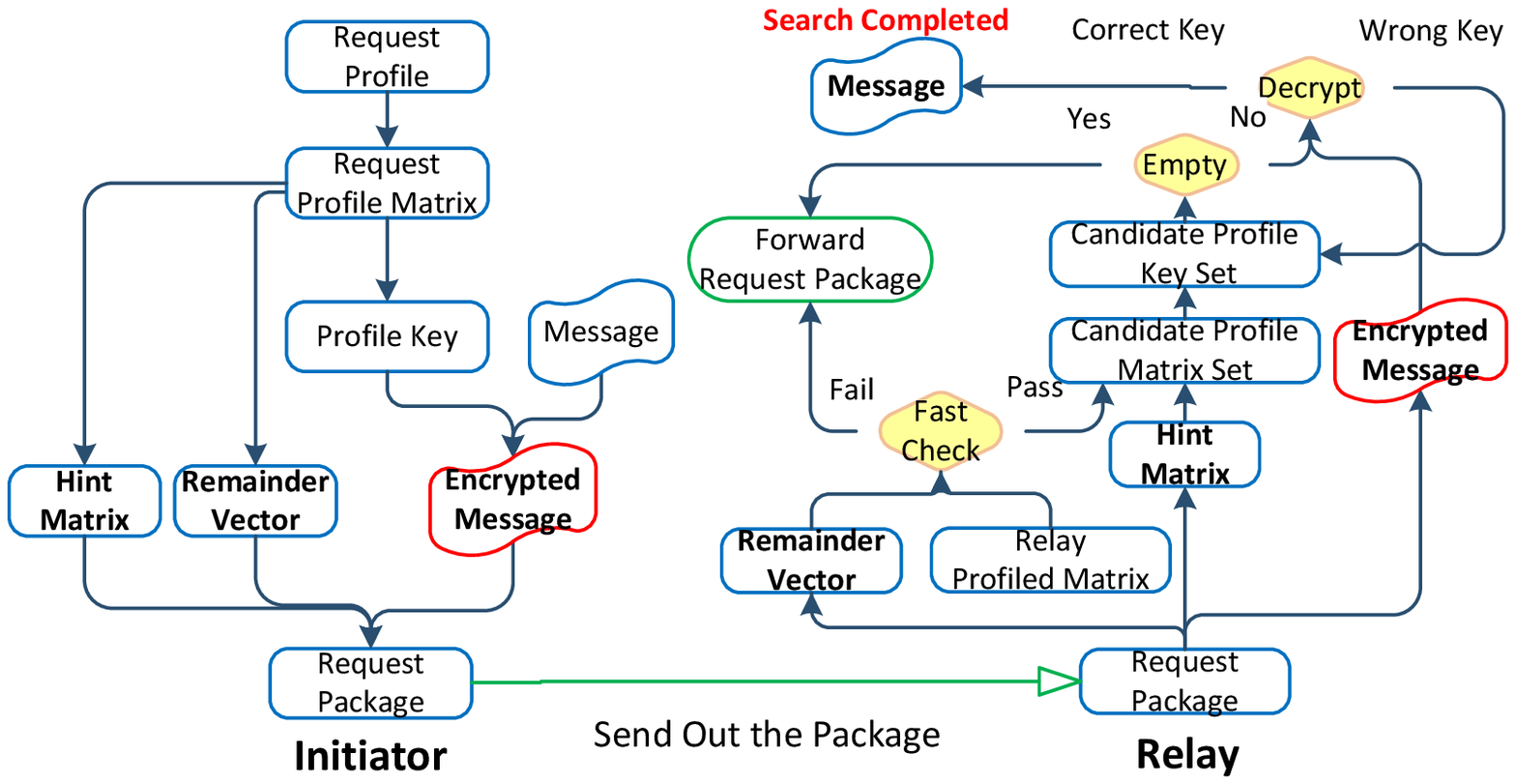}
\caption{The procedure to create a request package by the initiator and
handle a request package by some forwarder.}
\label{fig_mechanism}
\end{minipage}%
\hfill
\begin{minipage}[b]{0.45\linewidth}
\includegraphics[width=0.95\textwidth,height=1.7in]{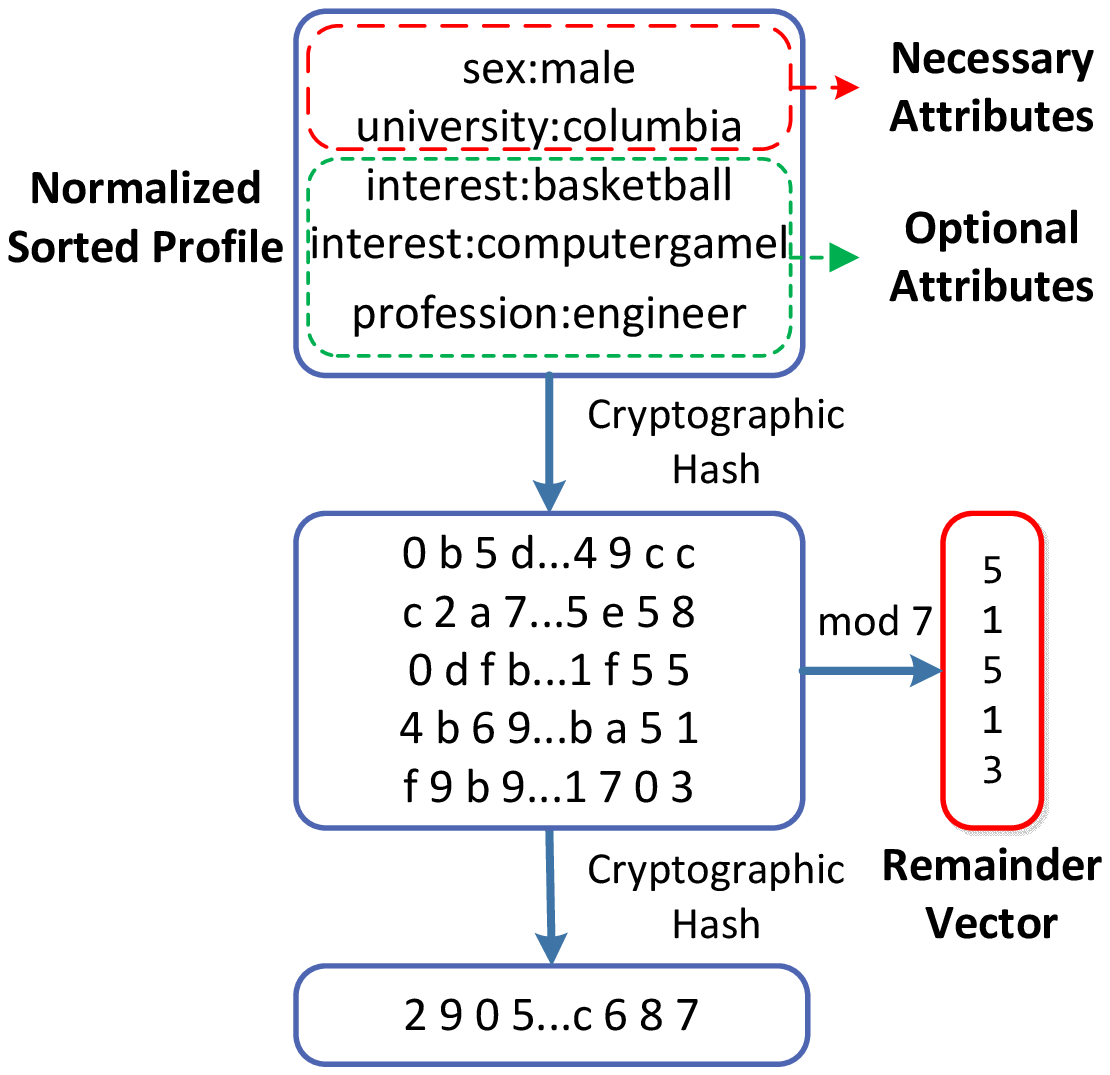}
\caption{The procedure to generate the \emph{profile key} and remainder vector with a sample user profile.}
\label{fig_matrix}
\end{minipage}
\end{figure*}



\subsection{Basic Mechanism}
\label{subsec:basic}

Observe that the intersection of the request profile
 and the match profile is a nature common secret shared by
 the initiator and the match user.
The main idea of our mechanism is to use the request profile as a key
 to encrypt a message.
Only a matching user, who shares the secret,
 can decrypt the message with his/her profile efficiently.

Fig.~\ref{fig_mechanism} illustrates our basic
 privacy preserving search and secure channel establishment mechanism.
Here, we first draw an outline of how the initiator creates a request package
 and how a relay user handles the request package.

\emph{The initiator} starts the process
 by creating a \emph{request profile} characterizing the matching user
 and a secret message for him/her.
 The request profile is a set of \emph{sorted} attributes.
Then he/she produces a \emph{request profile vector}
 by hashing the attributes of the request profile one by one.
A \emph{profile key} is generated based on the request profile vector using some publicly known hashing function.
The initiator encrypts the secret message with the profile key.
A \emph{remainder vector} of the profile vector is yield
 for fast exclusion by a large portion of unmatched persons.
To support a flexible fuzzy search requiring no perfect match,
 the initiator can define the \emph{necessary attributes},
 \emph{optional attributes} and the \emph{similarity threshold} of the matching profile.
And a \emph{hint matrix} is constructed from the request vector according to
 the similarity definition, which enables the matching person to recover
 the profile key.
In the end, the initiator packs
 the encrypted \emph{message}, the \emph{remainder vector} and the \emph{hint matrix}
 into a \emph{request package} and sends it out.
Note that the required profile vector will not  be sent out.

When a \emph{relay user} receives a request  from another user,
 he/she first processes a fast check of his/her own profile vector
 with the  remainder vector.
If no sub-vector of his/her profile vector fits the remainder vector,
 he/she  knows that he/she is not a matching user and
 will forward the request to other relay  users immediately.
Otherwise,  he/she is a \emph{candidate} target and
 will generate a \emph{candidate profile vector set}
 by some linear computation with his/her profile and the hint matrix.
Then a \emph{candidate profile key set} is obtained.
In the basic mechanism,
If any key of his/her candidate key set can decrypt the message correctly,
 he/she is an eligible person and the searching and private messaging complete.
Otherwise, he/she just forwards the request to other relay users.

\subsection{Profile Vector and Key Generation}
\label{subsec:profile-vector}

To protect the profile privacy and support a fuzzy search,
 a cryptographic hash (\eg SHA-256) of the attribute
 is adopted as the attribute equivalence criterion in this mechanism.
However, due to the avalanche effect,
 even a small change in the input will result in a mostly different hash.
Although consistent attribute name can be provided by the social networking service,
 the attribute fields or the tags are created by users.
So a profile normalization is necessary
 before the cryptographic hashing
 to ensure two attributes which are considered
 equivalent to yield the same hash value.
In our mechanism, we use some common techniques to normalize the users profile,
 including removing whitespace, punctuation, accent marks and diacritics,
 converting all letters to lower case, converting numbers (dates, currencies, etc.) into words,
  text canonicalization, expanding abbreviations,
 converting the plural words to singular form.
After the profile normalization,
 most inconsistences caused by spelling and typing are eliminated.
But the sematic equivalence between two different words are not in this paper's consideration.

Assume the cryptographic hash function is $\mathbf{H}$
 which yields $n$-bit length hash value.
With an \emph{sorted} normalized profile vector $A_k=[a_k^1, a_k^2, ... , a_k^m]^T$,
 a \emph{profile vector} $H_k$ is
 \begin{equation}
 H_k= \mathbf{H}(A_k)=[ h_k^1, h_k^2, ... , h_k^m]^T.
 \end{equation}
Here $h_k^i = \mathbf{H}(a_k^i)$, which is $n$ bit long.

A \emph{profile key} is created with $H_k$,
\begin{equation}
 K_k = \mathbf{H}(H_k).
 \end{equation}
Fig.~\ref{fig_matrix} shows the profile vector and key generation
 of an example profile. 
With the key of the required profile,
 the initiator encrypts the secret message
 using a symmetric encryption technique like Advanced Encryption Standard (AES).

\subsection{Remainder Vector and Hint Matrix}
\label{sec:remainder-vector}

So far, with the profile key,
 we have realized a naive private profile matching and secure channel establishment.
 The initiator sends out a message encrypted by the key of the required profile.
 Any person who receives it tries to decrypt the message
 with his/her own profile key.
 Only the exactly matching person will decrypt the message correctly,
 and he/she can construct a secure communication channel protected by the profile key with the initiator.
However, the naive mechanism has some flaws making it unpractical.
\begin{enumerate}
\item The search is not \emph{flexible}.
 The initiator cannot query any subset of other's profile.
 For example, he/she need to find a "student"
 studying "computer science" regardless of the "college".

\item An perfect matching is required and \emph{no fuzzy} search is supported.
 In most case, the initiator need only find some person with profile
 exceeding the required similarity threshold to the requested profile.

\item All participants must decrypt the message, although most of them hold wrong keys.
 It wastes the computation resource and increases the search delay.
\end{enumerate}

Improving the naive basic mechanism,
 our new mechanism allows the initiator to search a person with
 $m_t$ required attributes in a flexible way  $A_t=(N_t, O_t)$, as described in Section \ref{sec:model}.
In our mechanism we use a \emph{Remainder Vector}
 for fast excluding most unmatched users¡£
And a \emph{hint matrix} is designed to work with remainder vector
 to achieve efficient free-form fuzzy search.

\subsubsection{Remainder Vector}
\label{subsec:sample-vector}

Assume that there are $m_t$ attributes in the required profile,
$p$ is a prime integer larger than $m_t$.
A \emph{remainder vector} $R_k$ consists of the remainders
 of all the hashed attributes in the input profile vector divided by $p$,
 as illustrated in Fig.~\ref{fig_matrix}.
\begin{equation}
R_k = [h_k^1 \mod p, h_k^2 \mod p, ... , h_k^m \mod p]^T.
\end{equation}

Then the following theorem is straightforward.
\begin{theorem}\label{theorem1}
Given two attributes' hashes $h^i = \mathbf{H}(a^i)$
and $h^j = \mathbf{H}(a^j)$.
Remainder $r^i \equiv h^i \mod p$. Remainder $r^j \equiv h^j \mod p$.
If $r^i \neq r^j$, then $h^i \neq h^j$.
\end{theorem}

Assuming the required profile $A_t=[ a_t^1, a_t^2, ... , a_t^{m_t}]^T$ by an initiator,
the required profile vector is
$H_t=[h_t^1, h_t^2, ... , h_t^{m_t}]^T$.
Then the remainder vector of the required profile is $R_t = [r_t^1, r_t^2,...,r_t^{m_t}]^T$.
Assume that a relay user's profile vector is $H_k$.
With remainder vector, a relay user simply calculates
 $m_t$ \emph{candidate attribute subsets} $H_k(r_t^i)$ fitting each $r_t^i$.
Here
\begin{equation}
\forall h_k^x \in H_k(r_t^i): \ h_k^x \mod  p = r_t^i.
\end{equation}
\ie, attributes in $H_k(r_t^i)$ yields the same remainder $r_t^i$ when divided by $p$.

A combination
 of one element from each candidate attribute subset forms
 a profile vector of the relay user.
If the candidate attribute subset $H_k(r_t^i)$ is $\emptyset$,
 the corresponding element in the combination is \emph{unknown}
 and the relay user fails to meet the required attribute $a_t^i$
 according to Theorem~\ref{theorem1}.
The relay user is a \emph{candidate matching user} of the request if
 there exist at least one combination which is an ordered set, denoted by $H_c$, that satisfies the following:
\begin{enumerate}
\item The  $\alpha$ necessary attributes are all known, \ie
\begin{equation}
 H_k(r_t^i) \neq \emptyset, \forall i \leq \alpha;
\end{equation}
\item The number of unknown elements don't exceed $\gamma$, \ie
\begin{equation}
|\{H_k(r_t^i)|\alpha < i< m_t, \  H_k(r_t^i)=\emptyset\}| \leq \gamma;
\end{equation}
\item Since $H_t$ and $H_k$ are both sorted, the elements of $H_c$ should still keep
the order consistent with the relay user's profile vector $H_k$, \ie
\begin{equation}
\begin{aligned}
&\forall h_k^x \in H_c,h_k^y \in H_c.\\
&\ h_k^x \in H_k(r_t^i),  h_k^y \in H_k(r_t^j), i<j \Rightarrow  x<y.
\end{aligned}
\end{equation}
\end{enumerate}
We call $H_c$ a \emph{candidate profile vector}.

In a fuzzy search,
 during the fast checking procedure,
 if there is no candidate profile vector that can be constructed by the relay user's profile vector,
 then he/she is not a matching user and she/he can forward the package.
An ordinary relay user doesn't have many attributes, commonly dozens of attributes,
 so there won't be many candidate profile vectors.
Using the remainder vector, quick exclusions of a portion of unmatched users can be made.

\subsubsection{Hint Matrix}
\label{subsec:hint}

A \emph{hint matrix} is constructed to support a flexible fuzzy search.
It describes the linear constrain relationship
among the $\beta+\gamma$ optional attributes to help calculating
$\gamma$ unknown attributes from $\beta$ known attributes.
The hint matrix helps a matching user
 exceeding the similarity threshold to recover the required profile vector,
 so as to generate the correct profile key.
Note that when a perfect matching user is required, no hint matrix is needed.

The \emph{constrain matrix} with $\gamma$ rows and $\gamma+\beta$ columns is defined as:
\begin{equation}
C_{\gamma \times (\gamma + \beta)} = [I_{\gamma \times \gamma}, R_{\gamma \times \beta}].
\end{equation}
Here matrix $I$ is a $\gamma$ dimensional identity matrix,
 $R$ is a matrix of size $\gamma \times \beta$,
 each of its elements is a random nonzero integer.

Multiplying the constrain matrix to the optional attributes of the required profile vector,
the initiator gets a matrix $B$:
\begin{equation}
B = C \times [h_t^{\alpha+1}, h_t^{\alpha+2}, ... , h_t^r]^T
\end{equation}
Then the \emph{hint matrix} $M$ is defined as matrix $C$, followed by matrix $B$, \ie,
\begin{equation}
M = [C, B].
\end{equation}

When $\gamma > 0$, the initiator generates the hint matrix
and sends it with the encrypt message and the remainder vector.
In a fuzzy search,
 after the fast checking procedure by the remainder vector,
 if the relay user is a candidate matching user, he/she constructs a set of candidate profile vector $H_c$ with unknowns.

Guaranteed by the definition of the candidate profile vector, each $H_c$ has no more than $\gamma$ unknowns,
 and any unknown $h_c^i$, which is the $i$-th element of $H_c$, has $i>\alpha$.
Now, the unknowns of a candidate profile vector can be calculated
 by solving a system of linear equations:
\begin{equation}
C\times [h_c^{\alpha+1}, h_c^{\alpha+2}, ... , h_c^r]^T = B
\end{equation}
Equivalently, we have
\begin{equation}
[I, R] \times [h_c^{\alpha+1}, h_c^{\alpha+2}, ... , h_c^r]^T = B
\end{equation}
This system of equations has equal to or less than $\gamma$ unknowns.
It can be easily proved that, it has a unique solution.
With the solution, a complete candidate profile vector $H_c'$ is recovered.
For each $H_c'$, a candidate key $K_c = \mathbf{H}(H_c')$ can be generated.
If any of the relay user's candidate keys decrypts the message correctly,
 he/she is a matching user and gets the encrypted secret.
Else, he/she forwards the request to the next person.

%

\subsection{Location Attribute and Its Privacy Protection}
\label{subsec:location}

In localization enabled mobile social networks,
 a user usually searches matching users in vicinity.
In the existing systems, a user is required to
 provide his/her own current location information and
 desired search range.
However the user's accurate location will be exposed.
In our mechanism, location is considered as a dynamic attribute
 that will be updated while the user moves.
Accurate location is also the user's privacy.
To conduct location privacy preserved vicinity search,
 we design a scheme for location attribute matching based on
 lattice. Our location attribute matching scheme is compatible with
 our privacy preserved profile matching mechanism.
When generating the profile vector,
 we use lattice based hashing to hash a user's location and his/her vicinity region.
A private vicinity search can be easily conducted via our fuzzy search
 scheme with the help of hint matrix.

\subsubsection{Lattice based Location Hashing}

\begin{figure}[t!]
\centering
\includegraphics[width=0.3 \textwidth]{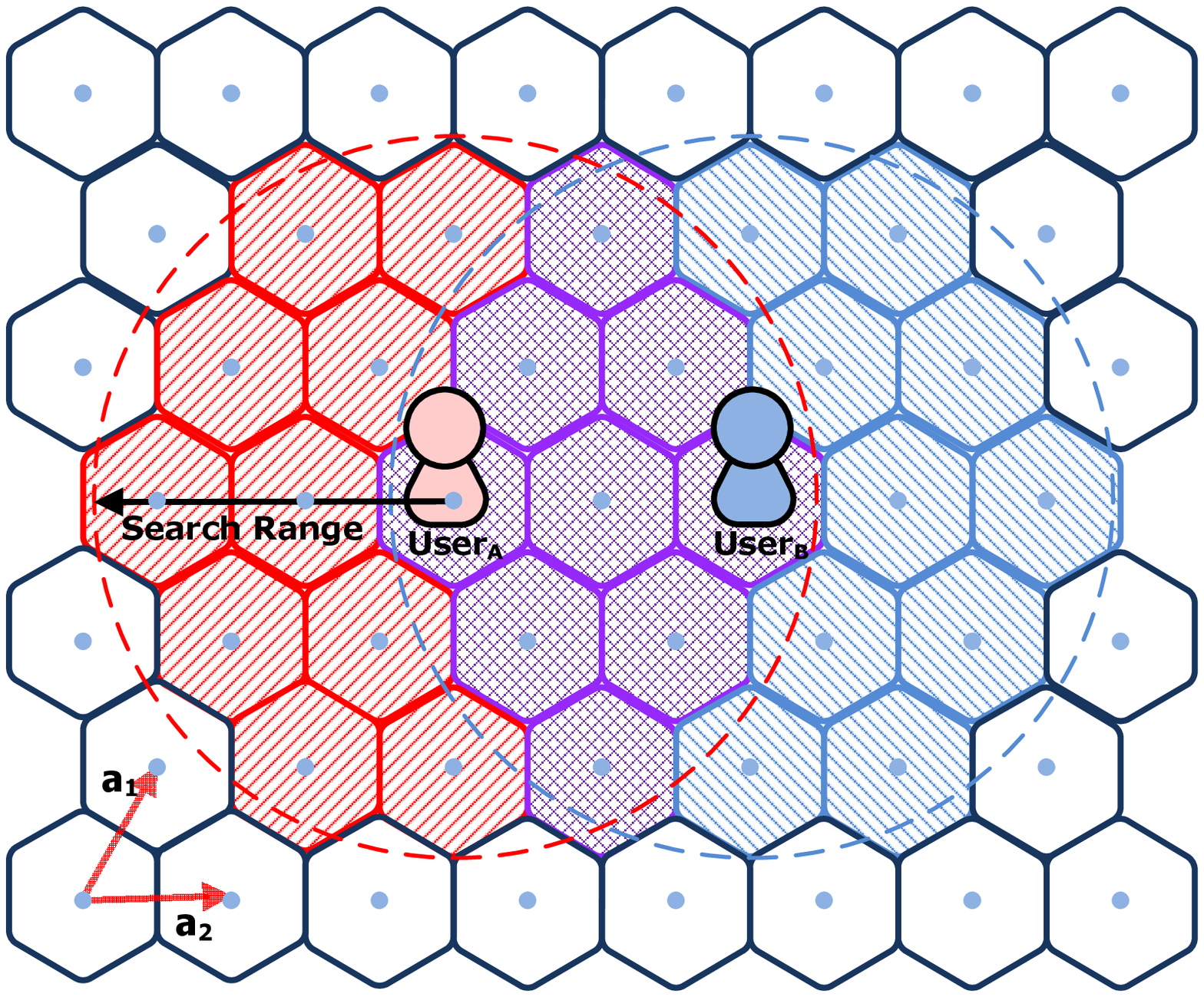}
\caption{Lattice-based location hash.}
\label{fig_lattice}
\end{figure}

We map the two-dimensional location to the hexagonal lattice.
The lattice point is a discrete set of the centers of all regular hexagons,
as the dots shown in Fig.~\ref{fig_lattice}.
The lattice is formally defined as:
\begin{equation}
\{x=u_1 a_1+u_2 a_2|(u_1,u_2)\in \mathbb{Z}^2\}
\end{equation}
Here $a_1,a_2$ are linearly independent primitive vectors which span the lattice.
Given the primitive vectors $a_1,a_2$,
 a point of the lattice is uniquely identified by the integer vector $u=(u_1, u_2)$.
There are infinite choices for $a_1, a_2$.
Let $d$ denote the shortest distance between lattice points,
for simplicity, we choose the primitive vectors as presented in Figure \ref{fig_lattice}:
\begin{equation}
a_1=(d,0);\
a_2=(\frac{1}{2}d, \frac{\sqrt{3}}{2}d).
\end{equation}

Defining a geography location as the origin point $O$
 and the scale of the lattice cell $d$, with the lattice definition,
 a location can be hashed to its nearest lattice point.
Any two locations hashed to the same lattice point are inside a single hexagonal lattice cell,
 and they are separated by a bounded distance $d$.

Given a user $v_k$'s current location $l_k(t)$ at time $t$ and his/her vicinity range $D$,
 his/her vicinity region can be hashed to a \emph{lattice point set}, $V_k(O,d,l_k(t),D)$,
 consisting of the center lattice point, i.e. the hash of $l_k(t)$, and other lattices points
 whose distances to the center lattice point are less than $D$.

\subsubsection{Location Privacy Preserved Vicinity Search}
Intuitively, given the distance bound to define vicinity,
 if two users are within each other's vicinity,
 the intersection of their vicinity regions will have a proportion no less than a threshold $\Theta$.
The initiator calculates his/her vicinity lattice point set $V_i(O,d,l_i(t),D)$
by defining $O$, $d$ and $D$, here $D$ indicates his/her search range.
Therefore, if a user $v_k$ is in his/her vicinity,
then $v_k$'s vicinity lattice point set $V_k(O,d,l_k(t),D)$ should satisfy the requirement:
\begin{equation}
\label{vinicity_threshold}
\theta_k = \frac{|V_i(O,d,l_i(t),D) \bigcap V_k(O,d,l_k(t),D)|}{|V_k(O,d,l_k(t),D)|} \geq  \Theta
\end{equation}
In the example in Fig.~\ref{fig_lattice}, user $v_A$'s vicinity range $D=3d$.
The lattice points within the red slashed hexagons constitute $V_A(O,d,l_A(t),D)$
and the lattice points within the blue back slashed hexagons constitute $V_B(O,d,l_B(t),D)$.
The purple grid region is the intersection of vicinity regions of $v_A$ and $v_B$.
In this case, the threshold $\Theta$ should be $\frac{9}{19}$.
In other words, if $V_B(O,d,l_B(t),D)$ contains no less than $9$ common lattice points of $V_A(O,d,l_A(t),D)$,
 he/she is in the vicinity of $v_A$.
 In practice, the initiator can adjust the parameter $d$ to
 change the cardinality of the vicinity lattice point set to a suitable size.

To conduct location privacy preserved vicinity search,
 the initiator won't send his/her vicinity lattice point set directly.
Using the sorted lattice points, he/she generates a dynamic profile key,
 a dynamic remainder vector and a dynamic hint matrix in the same way as processing other attributes.
So a vicinity search works as a fuzzy search with similarity threshold $\Theta$.
Only participants in his/her vicinity who has a certain amount of common lattice points with him/her
 can generate the correct dynamic profile key with the help of the dynamic remainder vector and hint matrix.

\subsubsection{Location Based Profile Matching}
Compared to static attributes like identity information,
 location is usually a temporal privacy.
With the \emph{dynamic key} generated from location information,
 we can improve the protection of a user's static attributes.
When constructing profile vector of a user,
 we can hash the concatenation of each static attribute and his/her current dynamic key
 instead of directly hash static attributes.
Since the dynamic key changes when the user updates his/her location,
 the hash values of the same static attribute are completely different for users at different locations.
It will greatly increase the difficulty for the malicious adversary to conduct dictionary profiling.
However, it won't bring much more computation for ordinary user because of quite limited lattice points
 in his/her vicinity range.

\subsection{Profile Matching Protocols}
\label{subsec:matching}

We are now ready to present our privacy preserving profile matching protocols. Here we present three different protocols that achieve
 different levels of privacy protection.

\subsubsection{Protocol 1}

\begin{algorithm}
\label{protocol1}
\textbf{Protocol 1:} Privacy Preserving Profile Matching
\begin{enumerate}
\item The initiator encrypts a random number $x$ and a public predefined confirmation information
in the secret message by the required profile key, say $E_{K_t}(confirmation,x)$. And he/she sends the request package out.

\item A candidate relay user can verify whether he/she decrypts the message correctly by the confirmation information.
If he/she does not match, he/she just forwards the request to the next user.
If he/she is matching,
 he/she can reply the request by encrypting the predefined ack information
 and a random number $y$ along with any other message (e.g. the intersection cardinality)
  by $x$, say $E_{x}(ack,y)$, and sends it back to the initiator.
\end{enumerate}
\end{algorithm}

The scheme of Protocol 1  is based on the basic mechanism.
Under Protocol 1,
a unmatched relay user doesn't know anything about the request.
The matching user knows the intersection of required profile and his/her own profile after step 1 in the HBC model.
A matching user can decide whether to reply the request according to the profile intersection.
The initiator doesn't know anything about any participant until he/she gets a reply.
With replies, he/she knows the matching users and even the most similar replier by the cardinality information.
Then he/she can start secure communication with a matching user protected by $x+y$ or with a group of matching user protected by $x$.
However, in malicious model, if the matching user has a dictionary of the attributes and the size of dictionary is not large enough unfortunately,
 he/she can learn the whole request profile by the recovered profile vector.

\subsubsection{Protocol 2}
To prevent malicious participants, we design Protocol 2, which is
 similar to Protocol 2, but it excludes the confirmation information from the encrypted message.

\begin{algorithm}
\label{protocol2}
\textbf{Protocol 2:} Privacy Preserving Profile Matching
\begin{enumerate}
\item The initiator encrypts a random number $x$ in the secret message
 without any confirmation information by the required profile key, say $E_{K_t}(x)$.
\item A candidate matching user cannot verify whether he/she decrypts the message correctly.
Let the candidate profile key set of a candidate user be $\{K_c^1, K_c^2,\dots, K_c^z\}$.
He/she decrypts the message in the request with each candidate profile key to get a set of numbers,
say $X=\{x_j|x_j=D_{K_c^j}(E_{K_t}(x))\}$.
Then he/she encrypts the predefined ack information and a random number $y$ by each $x^j$ as the key,
 and sends back the acknowledge set $\{E_{x^j}(ack,y)\}$ back to the initiator, for a public $ack$.
\item The initiator excludes the malicious replier whose response time exceeds the time window or
  the cardinality of reply set exceed the threshold.
\end{enumerate}
\end{algorithm}

Under Protocol 2, after the first round of communication,
 the participants won't known anything about the request in both HBC model and malicious model.
The initiator knows who are the matching users and even the most similar one according to the replies.
Then the initiator can start secure communication with a matching user protected by the key $x+y$
 or with a group of matching users protected by $x$.
In malicious model, if a participant has a dictionary of the attributes,
 he/she may construct a large candidate profile key set and send it to the initiator.
The main  difference between an ordinary user
 and a malicious user with a dictionary is the size of their attribute space.
An ordinary user with about dozens of attributes can make a quick reaction and reply a small size acknowledge set.
While it takes much longer for a malicious user due to a large number of candidate attribute combinations.
So the initiator can identify the malicious repliers by response time and the cardinality of reply set.

Using our profile matching mechanism, it is impossible to build an attribute dictionary in this social networking system.
If an adversary constructs the attribute dictionary from other sources, \eg, other similar social networking systems,
 the space of attributes are mostly very large, which makes the dictionary profiling infeasible.
Especially in the localizable social mobile social networks, the vast dynamic location attribute will greatly increase
 the difficulty of dictionary profiling.
However, there still may exist a special case that the attribute space is not large enough in some social networks.
In this case, in Protocol 1, the request profile may be exposed via dictionary profiling by malicious participants.
Although protocol 2 protects the request profile from any participants,
 a malicious initiator may learn the profile of unmatching repliers.

\subsubsection{Protocol 3}

To prevent  the dictionary profiling by malicious initiator, we improve Protocol 2 to Protocol 3
 which provides a user personal defined privacy protection.

\begin{definition}[Attribute Entropy]
For an attribute $a^i$ with $t^i$ values $\{x_j:j=1,\dots,t^i\}$.
$P(a^i=x_j)$ is the probability that the attribute $a^i$ of a user equals $x_j$.
The entropy of the attribute $a^i$ is
$S(a^i) = -\sum_{j=1}^{t^i}P(a^i=x_j)\log P(a^i=x_j)$.
\end{definition}

\begin{definition}[Profile Entropy]
The entropy of a profile $A_k$ is
$S(A_k) = \sum_{i=1}^{m_k}S(a^i)$.
\end{definition}


Intuitively, the larger the entropy of a profile,
 the more privacy information is contained in the profile.
A participant can determine his/her personal privacy protection level
 by given a acceptable profile entropy leakage upper limit $\varphi$.
Based on the user defined protection level,
 Protocol 3 is $\varphi$-entropy private for each user.

\begin{definition}[$\varphi$-Entropy Private]
A protocol is $\varphi$-entropy private
 when the entropy of possible privacy leakage is not greater than the upper limit $\varphi$:
$S(Leak(A_k)) \leq \varphi$.
\end{definition}

The parameter $\varphi$ is  decided by each user.
Here we suggest two options to determine $\varphi$.
\begin{enumerate}[(1)]
\item \emph{$K$-anonymity based}.
To prevent from being identified by disclosed attributes,
 the user will only send out messages protected by the profile key generated from the subset of attributes
 which at least $k$ users own the same subset.
Let a subset of $A_k$ be $A_k^s={a^1,\dots,a^l}$.
Suppose that the $t^i$ values of $a^i$ have equal probability,
 then there are $\frac{n}{\prod_{i=1}^{l}t^i}$ users are expected to own the same subset.
If the user require that $\frac{n}{\prod_{i=1}^{l}t^i} \geq k$,
 then $\log {\prod_{i=1}^{l}t^i} \leq \log\frac{n}{k}$, ie $S(A_k^s) \leq \log\frac{n}{k}$.
So the parameter $\varphi$ could be  $\log\frac{n}{k}$.
\item \emph{Sensitive attributes based}.
The user can determine the sensitive attributes which must not
  be disclosed according to the current context.
Let the set of sensitive attributes defined by user $v_k$ is $A_k^s$,
then $\varphi = \min(S(a^i))$, where $a^i \in A_k^s$.
\end{enumerate}

\begin{algorithm}
\textbf{Protocol 3:} Privacy Preserving Profile Matching
\begin{enumerate}
\item The initiator encrypts a random number $x$ in the secret message
 without any confirmation information by the required profile key, say $E_{K_t}(x)$.
\item A candidate matching user cannot verify whether he/she decrypts the message correctly.
He/she selects a set of candidate profile $\{A_c^1, A_c^2,\dots, A_c^z\}$ which
 satisfies that $S(\bigcup_{i=1}^z A_c^i) \leq \varphi$.
 And he/she generates the corresponding candidate profile keys $\{K_c^1, K_c^2,\dots, K_c^z\}$.
He/she decrypts the message in the request with each candidate profile key to get a set of numbers,
say $X=\{x_j|x_j=D_{K_c^j}(E_{K_t}(x))\}$.
Then he/she encrypts the predefined ack information and a random number $y$ by each $x^j$,
 and sends back the acknowledge set $\{E_{x^j}(ack,y)\}$ back to the initiator.
\item The initiator excludes the malicious replier whose response time exceeds the time window or
  the cardinality of reply set exceed the threshold and decrypts the replies with $x$.
  If he/she gets a correct ack information from a reply, the corresponding replier is matching.
\end{enumerate}
\end{algorithm}

Protocol 3 is private when the initiator is not malicious.
 and it is $\varphi$-private for each participant even
 when the initiator can conduct a dictionary profiling.

Note that, for all three protocols,
each request has a valid time.
 An expired request will be dropped.
And each user has a request frequency limit,
 all participants won't reply the request from the same user within a short time interval.

\subsection{Establishing Secure Communication Channel}
As presented in the profile matching protocols,
 the random number $x$ generated by the initiator and $y$ generated by a matching user have been exchanged secretly between them.
$x$ and $y$ is the communication key shared by a pair of matching users.
So the secure peer to peer communication channel can be constructed.
Our mechanism realizes key exchange between matching users which is resistant to the man in the middle attack.

Furthermore, our mechanism can be easily adopted to discover the community
 consisting of users with similar profile as the initiator and establish the group key $x$
 for secure intra-community communication.

\section{Security and Efficiency Analysis}
\label{sec:analysis}
In this section, we analyze the security and performance of our profile matching
 and secure communication channel establishing mechanism.

\subsection{Security and Privacy Analysis}

\subsubsection{Profile Privacy}
During the generation of the profile key,
 we don't use the hash value of the static attribute directly.
Instead, we use the hash value (with hash function SHA-256) of the combination of the static attribute
 and the dynamic attribute (\ie location),
 which greatly increases the protection of the static attribute.
We use AES with 256 bit key as the encryption method.
The 256-bit profile key is used as the secret key to encrypt the message by AES.
Only the encrypted message will be transmitted,
 and no attribute hash value will be transmitted.
No one can obtain other user's attribute hash.
Therefore no one can build a dictionary of hash values of attributes
 via this social networking system.
 To acquire the profile information of the initiator or other participants
 the attacker need to decrypt the request/reply message correctly and confirm the correctness.
This is extremely difficult due to the choice of SHA-256 and 256-bit-key AES.

\begin{table}[tbhp]
\caption{The privacy protection level of our protocols in HBC model.
$v_I$ is the initiator, $v_M$ is a matching user and $v_U$ is a unmatching user.
$A_I$, $A_M$ and $A_U$ are their corresponding profiles.}
\label{table:security1}
\centering
\begin{tabular}{|c|c|c|c|c|}
\hline
$PPL$ & $(A_I,v_M)$ & $(A_I,v_U)$ & $(A_M,v_I)$ & $(A_U,v_I)$ \\
\hline
Protocol 1 & 1 & 3 & 2 & 3\\
\hline
Protocol 2 & 3 & 3 & 2 & 3\\
\hline
Protocol 3 & 3 & 3 & 2 & 3\\
\hline
PSI  & 3 & 3 & 1 & 1\\
\hline
PCSI  & 3 & 3 & $|A_I\cap v_U|$ & $|A_I\cap v_U|$\\
\hline
\end{tabular}
\end{table}

\begin{table*}[tb]
\caption{The privacy protection level of our protocols in malicious model with small dictionary.
$v_I$ is the initiator, $v_M$ is a matching user and $v_U$ is a unmatching user.
$A_I$, $A_M$ and $A_U$ are their corresponding profiles.
$v_I'$ is a malicious initiator with a profile dictionary.
$v_P'$ is a malicious participant with a profile dictionary eavesdropping the communication.}
\label{table:security2}
\centering
\begin{tabular}{|c|c|c|c|c|c|}
\hline
$PPL$ & $(A_I,v_P')$ & $(A_M,v_I')$ & $(A_M,v_P')$ & $(A_U,v_I')$ & $(A_U,v_P')$ \\
\hline
Protocol 1 & 0 & 2 & 2 & 3 & 3\\
\hline
Protocol 2 & 3 & 2& 3 & \tabincell{c}{3 (noncandidate) \\$A_c$ (candidate) } & 3\\
\hline
Protocol 3 & 3 & $\varphi$-entropy & 3 & \tabincell{c}{3 (noncandidate) \\$\varphi$-entropy (candidate)} & 3 \\
\hline
\end{tabular}
\end{table*}

In the honest-but-curious (HBC) model each user only knows his/her own attributes.
Therefore, only users owning matching attributes
 can generate the same profile key and decrypt each other's messages correctly.
Unmatched user cannot obtain any information from the encrypted message.
Table \ref{table:security1} presents the protection level of our three protocols during the matching procedure in HBC model.
Compared to the existing PSI and PCSI approaches,
 our protocols provide Level 2 privacy protection for matching users
 and don't leak any information to unmatching users..
After profile matching,
 under Protocol 1, the initiator knows nothing until the matching user replies;
 under Protocol 2 and 3, a user won't learn he/she is matching until the initiator contacts him/her.

In the malicious model,
for the adversary who intends to learn the attributes hash to build a dictionary,
 it is impossible in our social networking system, as a result of that no attribute
 information is transmitted in any data packets.
But we still consider the unlikely case that an adversary
 constructs an attribute dictionary from other sources,
 \eg, other similar social networking systems.
We suppose that the adversary can eavesdrop all communication.

In most cases, the cardinality $m$ of the dictionary is very large,
 which makes the dictionary profiling difficult.
With a remainder vector,
 it takes an adversary $(\frac{m}{p})^{m_t}$ guesses to compromise a user's profile with $m_t$ attributes.
Here $p$ is a small prime number like $11$.
In tencent weibo, we found that $m\simeq 2^{20}$ and the average attribute number of each user is $6$.
When the adversary tries to guess a user's profile by brute force,
it will take about $2^{100}$ guesses.
If considering keywords of a user, $m$ is even larger.
Especially in localizable mobile social networks,
 the vast dynamic location attribute will greatly increase the attribute set.
The dictionary profiling gets more infeasible.

However, the worst case may still exist that
 the adversary obtains the attribute dictionary and the dictionary size is not large enough.
This kind of adversary also compromises other PSI based approaches.
Table \ref{table:security2} shows the protection level of our protocols in this worst case.
In this case, our Protocol 1 cannot protect the initiator's privacy from the dictionary profiling.
 But it provides Level 2 privacy protection for replying matching users and unconditional Level 3 privacy for unmatching users.
Protocol 2 provides unconditional Level 3 privacy for the initiator.
 It also provides unconditional Level 3 privacy for all participants against any other person except the initiator.
 Only if the initiator is an adversary with the dictionary, he/she may compromise the candidate user's privacy.
Although candidate users are just a small potion of users, to improve Protocol 2 to protect their privacy,
 we design Protocol 3 which still provides unconditional Level 3 privacy for the initiator and incardinate users,
   and Level 3 privacy for the candidate users against any other person except the initiator.
Moreover, it provides personal defined $\varphi$-entropy private for all candidate users against an malicious initiator.

\subsubsection{Communication Security}
Our protocols realize secure key exchange between matching users
 based on their common attributes.
The shared secret key is protected by the profile key,
 only the user who owns the matching attributes can generate the same profile keys.
As the security analysis of our protocols,
 Protocol 2 and 3 can construct a secure communication between a matching user and the initiator
 against any other adversary in both HBC and malicious model.
So our secure communication channel establishment between matching users thwarts the MITM attack.

\subsubsection{Verifiability}
In majority of existing profiling matching approaches,
  only one party learns the result and tells the other party.
Hence, one party can lie about the result.
There lacks a feasible way for the other party to verify the result.
Our protocols are verifiable and resists cheating.
In Protocol 1, after the matching, only the matching user learns the result
 and he/she contacts the initiator by relying $E_x(ack,y)$.
In the HBC model, an unmatching user cannot obtain the correct $x$ encrypted by the request profile key,
 so he/she cannot cheat the initiator to pretend to be matching.
In Protocol 2 and 3, similarly, only the matching user can get the correct $x$
Consequently, the initiator can only obtain the correct $y$ of matching user.
So both the initiator and participants cannot cheat each other.
In the malicious model with a  dictionary,
 it takes an adversary much longer time to guess the correct key.
Hence a nonmalicious user can distinguish the adversary by his/her reply delay.
So no cheating can be conducted successfully with our protocols.


\subsection{Performance Analysis}

\subsubsection{Computational Cost}
For an initiator, it takes $m_t\log m_t$ operations for  sorting attributes,
 $m_t+1$ hashing operations for profile key generation and $m_t$ modulo operations for remainder vector generation.
$\gamma(\gamma+\beta)$ operations are needed to calculate the hint matrix if the required similarity $\theta < 100\%$.
In addition, one symmetric encryption is needed with the profile key.

For a participant $v_k$, it takes $m_k\log m_k$ operations for  sorting its attributes,
 $m_k$ hashing for profile vector generation and $m_k$ modulo operations for remainder calculation.
After fast checking by remainder vector, if a participant $v_k$ is not a candidate user, no more computation is need.
If $v_k$ is a candidate user, let the number of his candidate profile vector be $\kappa_k$.
It takes this user $\kappa_k$ hashing to generate candidate profile keys.
If there is a hint matrix, user $v_k$ need to solve $\kappa_k$ $\gamma(\gamma+\beta)$-dimension linear equation systems.
The computation cost is $\kappa_k m_t^3$.
In the end, $\kappa_k$ symmetric decryptions with the profile key.
Note that the expected $\kappa_k$ is $\varepsilon(\kappa_k) = \binom{m_k}{\alpha+\beta} \times \left(\frac{1}{p}\right)^{\alpha+\beta}$.
For example, in Tencent Weibo the average attribute number is $6$ and the maximum number is $20$.
 Then if $\alpha+\beta=6$, even a large $m_k=20$ and small prime number $p=11$ result in a very small $\varepsilon(\kappa_k)=0.02$.
So it takes small computation cost even for a candidate user.
We can show that the expected candidate users is only a small portion of all users and the portion decreases greatly with the increase of $m_t$ and $p$.
It may be considered that larger $p$ will weaken the security due to the decreased difficulty of dictionary profiling.
Our testing and analysis show that even a small $p$, \eg, $p=11$, can significantly reduce the number of candidate users.
So an initiator can choose a proper $p$ to efficiently control the amount of candidate users
 as well as achieve the secure protection of profile privacy.

In implementation, we use SHA-256 as the hashing method and AES as the symmetric encryption method.
The data processing performance of SHA-256 is $111$MB/s for a single-threaded implementation
 on an Intel $1.83$GHz processor in 32-bit operation system.
Furthermore, for all users the sorting and hashing results are calculated once and used repetitively
 until the attributes are updated, \eg, the location changes.
AES performs well on a wide variety of hardware, from 8-bit smart cards to high-performance computers.
The throughput of AES is about $60$MB/s on a 1.7GHz processor and about $400$MB/s on Intel i5/i7 CPUs.
Compared to the existing approaches, we don't use an asymmetric-key cryptosystem
 and the remainder vector can significantly reduce the computation of unmatching users.
In all, our protocol are computationally efficient.

\subsubsection{Communication Cost}

In our protocols there isn't any pre-operation for key exchange or secure channel construction.
To conduct profile matching with all users,
 the initiator only need one broadcast to send the request to all participants.
The request consists of a $32m_t$ bit remainder vector and a  $256$ bit encrypted message.
 If the required similarity $\theta < 100\%$, there is also a $32\gamma(\gamma+\beta)+256\gamma$ bit hint matrix.
So the size of the request message is at most $(1-\theta)32m_t^2 + (288-256\theta)m_t + 256$ bits.
For example, the user of Tencent Weibo has 6 tags in average and 20 tags at most.
To search a $60\%$ similar user, the request is about $190$B in average and $1$KB at most. 
In Protocol 1, only the matching user will reply the request.
 So the transmission cost of Protocol 1 is one broadcast and $O(1)$ unicasts.
In Protocol 2, only the candidate matching user will reply the request.
 So the transmission cost of Protocol 2 is one broadcast and $O(n*(\frac{1}{p})^{m_t\theta})$ unicasts.
 For example, when $p=11$, $m_t=6$, $\theta=0.6$, there are only about $\frac{1}{5610}$ fraction of users will reply.
In Protocol 3, the communication cost of reply is even smaller than Protocol 2 because of the personal privacy setting.
Note that the reply in all  three protocols is only $32$Byte.
So the communication cost of our protocols is quite small.
This makes our protocols  suitable for wireless communication environment, for example the mobile social networks.

\subsubsection{Further Comparison With Related Work}

In this section, we mainly compare the computation cost and communication cost
 between our protocols and PSI and PCSI based approaches.
The computation time comparison with dot  product based on approach presented in \cite{dong2011secure}.

First we define the basic operations of other asymmetric cryptosystem based approaches:
\begin{itemize}
\item $\boldmath{M}_1$: 24-bit modular multiplication;
\item $\boldmath{M}_2$: 1024-bit modular multiplication;
\item $\boldmath{M}_3$: 2048-bit modular multiplication;
\item $\boldmath{E}_2$: 1024-bit exponentiation;
\item $\boldmath{E}_3$: 2048-bit exponentiation.
\end{itemize}
Then we define the basic operation in our protocol:
\begin{itemize}
\item $\mathcal {H}$: one SHA-256 hashing operation of an attribute;
\item $\mathcal {M}$: is the operation that the 256-bit hash of an attribute modulo the small prime $p$;
\item $\mathcal {E}$: AES-256 encryption;
\item $\mathcal {D}$: AES-256 decryption.
\end{itemize}

Table \ref{table:compare1} presents the computation and communication cost
 of related work and our protocol.
Because the SHA-256, modulo, and AES-256 operations in our protocol are much cheaper than
 the 1024-bit and 2048-bit modular multiplication and exponentiation,
 our protocol costs much less computation  than asymmetric-key based schemes.
The transmitted data size are significantly reduced because basically only one encrypted 256-bit message
 and one $32*m_t$ bit remainder vector need to be transmitted.
Furthermore, the remainder vector eliminates the reply from most unmatching users.
So the total transmission time is also reduced.

\begin{table*}
\centering
\caption{Comparison of efficiency with existing scheme.$q=256$}
\begin{tabular}{|c|c|c|c|c|c|}
\hline
& Party & FNP\cite{freedman2004efficient} & FC10\cite{de2010practical} &  Advanced\cite{li2011findu} & Protocol 1 \\
\hline
Computation & \tabincell{c}{$P_1$\\$P_k$} & \tabincell{c}{$(2m_t+m_kn)\boldmath{E}_3$ \\ $m_k\log m_t\boldmath{E}_3$} &
\tabincell{c}{$(2.5m_tn)\boldmath{M}_2$ \\ $(m_t+m_k)\boldmath{E}_2$} &
\tabincell{c}{$(3m_tn)\boldmath{E}_3$ \\ $2m_t\boldmath{E}_3$}&
\tabincell{c}{$(m_t+1)\mathcal{H}+m_t\mathcal{M} + \mathcal{E}$ \\ $m_k\mathcal{H} + m_k\mathcal {M}$ (noncandidate)\\
$\kappa_k\gamma^2(\gamma+\beta)+  (m_k+ \kappa_k)\mathcal{H}$ \\ $+ m_k\mathcal {M} + \kappa_k\mathcal {D}$ (candidate)}\\
\hline
\tabincell{c}{Communication\\bits} & All & $8q(m_t+m_kn)$ & $4qn(3m_t+m_k)$ & \tabincell{c}{$24[m_tm_kn$\\$+tn(8m_t+2m_k+12m_tt)]$\\$+16qm_tn$} &
\tabincell{c}{$(1-\theta)32m_t^2 + (288-q\theta)m_t$\\$ + q + qn*(\frac{1}{p})^{m_t\theta}$}\\
\hline
\tabincell{c}{Communication\\Transimission} & All &  \tabincell{c}{$1$ broadcast\\$n$ unicasts} & $2n$ unicasts & $5n$ unicasts & \tabincell{c}{$1$ broadcast\\$n*(\frac{1}{p})^{m_t\theta}$ unicasts} \\
\hline
\end{tabular}
\label{table:compare1}
\end{table*}

\section{Evaluation Using Real Data}
\label{sec:evaluation}

\subsection{Real Social Networking System Analysis}

Our evaluations are based on the profile data of Tencent Weibo \cite{tencent}.
Tencent Weibo is one of the largest micro-blogging websites in China,
 which is a platform for building friendship and sharing interests online.
This dataset has $2.32$ million users personal profiles,
 including the following information: the year of birth, the gender, the tags and keywords.
Tags are selected by users to represent their interests.
Keywords are extracted from the tweet/retweet/comment of a user,
 and can be used as features to better represent the user.

In the Tencent Weibo dataset, the cardinality of the tag set is $560419$ and
 the cardinality of the keywords set if $713747$.
Each user has $6$ tags in average and $20$ tags at most to represent his/her interest.
On average, $7$ keywords extracted for a user.
 The maximum number of keywords  of a user is $129$.
So when the adversary tries to guess a user's profile with $6$ tags by brute force,
it will take him/her about $10^{30}$ guesses.
Moreover, the large attribute space makes the vector-based matching approaches impractical.

\begin{figure}[htb]
\begin{minipage}[b]{0.45\linewidth}
\includegraphics[width=0.9 \textwidth]{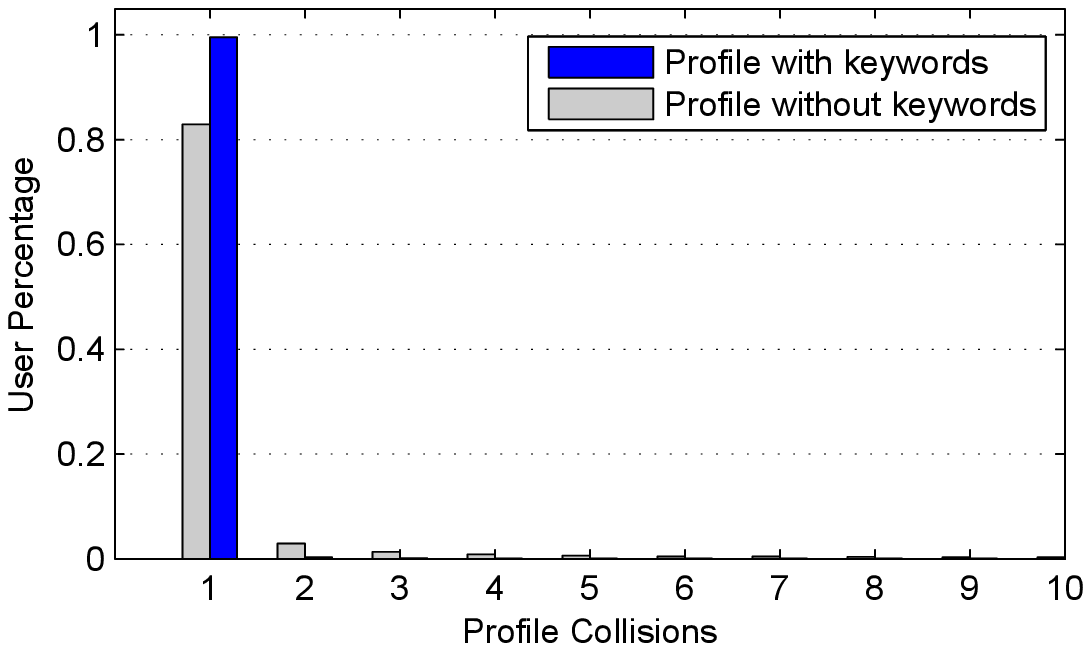}
\caption{Profile uniqueness and collisions of users.}
\label{fig_profile_collision}
\end{minipage}%
\hfill
\begin{minipage}[b]{0.45\linewidth}
\includegraphics[width=0.9\textwidth]{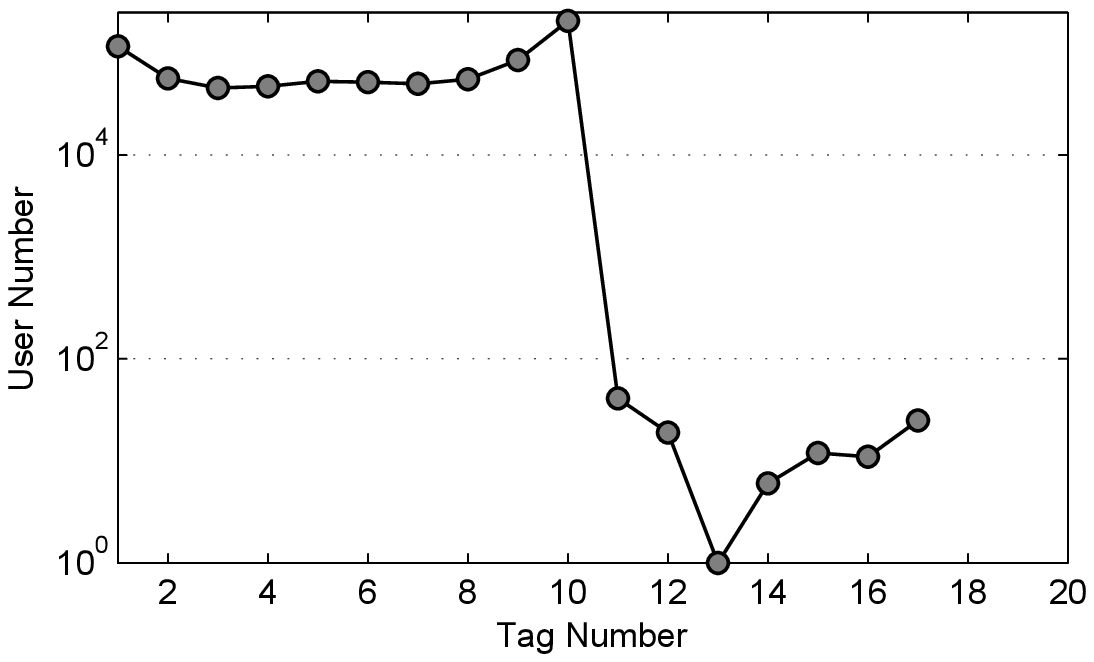}
 \caption{Users' attribute number distribution.}
\label{fig_tag_num}
\end{minipage}
\end{figure}

There is a question that, would many users own the same profile?
Each profile is a combination of several attributes.
When more than one users have the same profile,
we say there are collisions for this profile.
Figure \ref{fig_profile_collision} shows that both in Tencent
and Facebook more than $90\%$ users have unique profiles.




\subsection{Computation Performance}

We then exam the computation performance of our protocols on mobile devices and PC.
The mobile phone is HTC G17 with $1228Hz$ CPU, 1GB ROM and 1GB RAM.
The laptop is Think Pad X1 with i7 $2.7$GHz CPU and 4GB RAM.
Table \ref{table:optime} shows the mean execution time of basic computation.
And we evaluate our protocol on the Tencent Weibo dataset,
Table \ref{table:time} presents the breakdown of time cost for our protocol.
As an example, for a user with $6$ attributes, the time need to generate a request is only about
 $3.9 \times 10^{-2}$ ms on laptop and $1.3$ ms on mobile phone.
On  average it takes a non-candidate user about $3.9 \times 10^{-2}$ ms on laptop
 and $0.63$ ms on phone to process the request.
A little more extra computation is required for a candidate user,
 and the computation cost is about $4 \times 10^{-2}$ ms on laptop and $7$ ms on phone for each candidate key.

The time cost of all the operations in our protocols are quite small
 compared with the computation time of existing asymmetric
 cryptosystem based approaches, \eg the evaluation result of \cite{dong2011secure}.
Table \ref{table:compare2} shows a typical scenario in a mobile social network with 100 users.
The  numbers of attributes are chosen based on the analysis of Tencent Weibo.
With the numerical comparison, it is clearly that our protocol is efficient in both computation and communication.

\begin{table}[hptb]
\caption{Mean computation time for our basic operation.(ms)}
\label{table:optime}
\centering
\begin{tabular}{|c|c|c|c|}
\hline
& SHA-256 & Mod $p$ & AES Enc \\
\hline
Laptop & $1.2 \times 10^{-3}$ & $3.1 \times 10^{-4}$ & $8.7 \times 10^{-4}$ \\
\hline
Phone & $4.8 \times 10^{-2}$ & $5.7 \times 10^{-2}$ & $2.1 \times 10^{-2}$ \\
\hline
& Multiply-256 & Compare-256 & AES Dec\\
\hline
Laptop & $1.4 \times 10^{-4}$ & $1.0 \times 10^{-5}$ & $9.6 \times 10^{-4}$ \\
\hline
Phone & $3.2 \times 10^{-2}$  & $1.0 \times 10^{-3}$ & $2.5 \times 10^{-2}$\\
\hline
\end{tabular}
\end{table}

\begin{table}[hptb]
\caption{Mean computation time for basic operations for asymmetric cryptosystem based scheme.(ms)}
\label{table:optime2}
\centering
\begin{tabular}{|c|c|c|c|c|}

\hline
& 1024-exp & 2048-exp & 1024-mul & 2048-mul \\
\hline
Laptop & $17$ & $120$ & $2.3 \times 10^{-2}$ & $1 \times 10^{-1}$\\
\hline
Phone & $34$ & $197$ & $1.5 \times 10^{-1}$  &$2.4 \times 10^{-1}$ \\
\hline
\end{tabular}
\end{table}

\begin{table}[hptb]
\caption{Decomposed computation time of our protocols based on the Tencent Weibo dataset.(ms)}
\label{table:time}
\centering
\begin{tabular}{|c|c|c|c|}
\hline
\multicolumn{4}{|c|}{Laptop}\\
\hline
& Mean & Min & Max\\
\hline
MatrixGen & $7.2 \times 10^{-3}$ & $1.0 \times 10^{-3}$ & $2.4 \times 10^{-2}$\\
\hline
KeyGen & $8.1 \times 10^{-3}$ & $2.3 \times 10^{-3}$ & $2.5 \times 10^{-2}$\\
\hline
RemainderGen & $1.9 \times 10^{-3}$ & $3.1 \times 10^{-4}$ & $6.2 \times 10^{-3}$\\
\hline
HintGen &  $4.7 \times 10^{-3}$ & $2.8 \times 10^{-4}$ & $5.6 \times 10^{-2}$\\
\hline
HintSolve & $3 \times 10^{-2}$ & $1.1 \times 10^{-3}$ & $1.1$\\
\hline
\multicolumn{4}{|c|}{Phone}\\
\hline
& Mean & Min & Max\\
\hline
MatrixGen & $2.6 \times 10^{-1}$ & $4.4 \times 10^{-2}$  & $8.9 \times 10^{-1}$ \\
\hline
KeyGen & $6.3 \times 10^{-2}$ & $4.8 \times 10^{-2}$ & $1.4 \times 10^{-1}$\\
\hline
RemainderGen & $3.4 \times 10^{-1}$ & $5.7 \times 10^{-2}$ & 1.14\\
\hline
HintGen &  $1.2$ & $1.4 \times 10^{-1}$ & $12$  \\
\hline
HintSolve & $6.9$ & $2.6 \times 10^{-1}$ & 250\\
\hline
\end{tabular}
\end{table}

 \begin{table*}[tb]
\label{table:compare2}
\centering
\caption{Comparison of efficiency with existing scheme in typical scenario.
$m_t=m_k=6$, $\gamma=\beta=3$, $p=11$, $n=100$, $t=4$.}
\begin{tabular}{|c|c|c|c|c|c|}
\hline
& Party & FNP\cite{freedman2004efficient} & FC10\cite{de2010practical} &  Advanced\cite{li2011findu} & Protocol 1 \\
\hline
Computation & \tabincell{c}{$P_1$\\$P_k$} & \tabincell{c}{$612\boldmath{E}_3$ \\ 5$\boldmath{E}_3$} &
\tabincell{c}{$1500\boldmath{M}_2$ \\ $12\boldmath{E}_2$} &
\tabincell{c}{$1800\boldmath{E}_3$ \\ $12\boldmath{E}_3$}&
\tabincell{c}{$7\mathcal{H}+6\mathcal{M} + \mathcal{E}$($P_1$) \\ $6\mathcal{H} + 6\mathcal {M}$ (noncandidate)\\
 $27\kappa_k+ (6+ \kappa_k)\mathcal{H} + 6\mathcal {M} + \kappa_k\mathcal {D}$ (candidate)}\\
\hline
Computation(ms) & \tabincell{c}{$P_1$\\$P_k$} & \tabincell{c}{$73440$ \\ $600 $}&
\tabincell{c}{$34.5$ \\ $204$} &
\tabincell{c}{$216000$ \\ $1440$}&
\tabincell{c}{$1.1 \times 10^{-2}$($P_1$) \\ $3.1 \times 10^{-3}$(noncandidate)\\ $6\kappa_k \times 10^{-3}+ 9.1 \times 10^{-3}$ (candidate)}\\
\hline
Communication(KB) & All & $151$ & $300$ & $704$ &$0.22$\\
\hline
\tabincell{c}{Communication\\Transimission} & All&  \tabincell{c}{$1$ broadcast\\$100$ unicasts} & $200$ unicasts & $500$ unicasts & \tabincell{c}{$1$ broadcast\\ number of candidates unicasts} \\
\hline
\end{tabular}
\end{table*}

\subsection{Performance Evaluations}

Based on the user attributes of Tencent Weibo,
 we evaluate the efficiency of our protocols.
Two typical situations are taken into consideration.
In the first case, all users have equal size of attributes which is similar
 to the vector based scheme.
We use the attribute data of all $52248$ users with $6$ attributes.
In the second case, we randomly sample $1000$ users from all users to conduct profile matching.

\begin{figure}[t!]
\begin{tabular}{cc}
\includegraphics[width=0.22\textwidth]{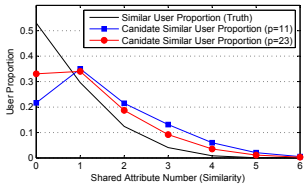} &
\includegraphics[width=0.22\textwidth]{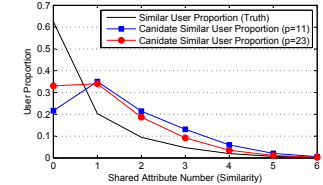}\\
(a) users with 6 attributes & (b) diverse numb. of attributes
\end{tabular}
\caption{Candidate user proportion with different similarity and prime number. }
\label{fig_candidate_tag6}
\label{fig_candidate_sample}
\end{figure}

In both cases, we exam the similarity between all pairs of users as the ground truth.
Then we run our profile matching protocols at different similarity levels.
Figure \ref{fig_candidate_tag6}  
shows the number of candidate users of our protocol change with similarity requirement and the prime number $p$.
The result shows that in both cases, the number of candidate users  approaches the number of true matching users with increasing $p$.
And a small $p$ can achieve small size of candidate users and significantly reduce unwanted computation and communication cost for unmatching users.

\begin{figure}[t!]
\begin{tabular}{cc}
\includegraphics[width=0.22\textwidth]{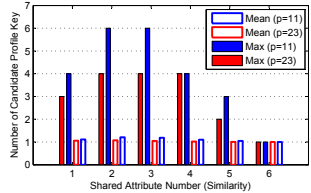} &
\includegraphics[width=0.22\textwidth]{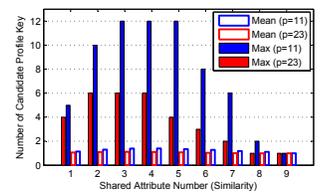} \\
(a) users with 6 attributes & (b) diverse num. of attributes
\end{tabular}
 \caption{Size of candidate profile key set with different similarities.}
\label{fig_key_tag6}
\label{fig_key_sample}
\end{figure}

There is a worry that, the candidate profile key set may be very large for candidate users.
Figure \ref{fig_key_tag6} 
presents the number of candidate profile keys
 during the matching with different similarity level and prime $p$.
The result shows that in real social networking system like Tencent Weibo,
 the candidate key set is small enough to achieve efficient computation for candidate users.

\section{Related work}
\label{sec:review}
\subsection{Privacy Preserving Friending}

Most previous private matching work are based on the
 secure multi-party computation (SMC) \cite{yao1982protocols}.
There are two mainstreams of approaches to solve the private profile-based friending problem.
The first category is based on private set intersection (PSI) and private cardinality of set intersection  (PCSI).
Early work in this category  mainly address the private set operation problem in database research, \eg \cite{agrawal2003information,freedman2004efficient,kissner2005privacy,ye2008distributed,sathya2009multi,de2010practical}.
Work like \cite{qi2008efficient} studied the problem of private $k$ nearest neighbor search.
Later on some work treat a user's profile as multiple attributes chosen from a public set of attributes
 and provide well-designed protocols to privately match users' profiles based on PSI and PCSI, \cite{von2008veneta,li2011findu}.
The second category is based on private vector dot product \cite{ioannidis2002secure}.
\cite{dong2011secure} considers a user's profile as vector which represents his/her social coordinate,
 and the social proximity between two uses as the matching metric.
 It calculates the metric by private dot product.
A trusted central server is requited to precompute users social coordinates and generate certifications and keys.
\cite{zhang2012fine} improves these work with a fine-grained private matching
 by associating a user-specific numerical value with every attribute to indicate
 the level of interest.
However, most these approaches lacks a specific definition of matching user.
For example, in the PSI based schemes, any user can learn the profile intersection
 with any other user despite the similarity between them.
The PCSI and dot product based approaches cannot support a precise specific profile matchings.

These protocols often rely on public-key cryptosystem and/or homomorphic encryption,
 which results in expensive computation cost and usually requires a trusted third party.
 Even non-matching users involve in the expensive computation.
Multiple rounds of interactions are required to perform the private profile
 matching between each pair of parties.
They all need preset procedure, \eg exchanging public keys before matching,
 precomputing vector \cite{dong2011secure}, establishing secure communication channel and share secret \cite{li2011findu}.
Furthermore, these protocols are unverifiable.

\subsection{Establishing Secure Channel}

Secure communication channel construction is very important
 in practical private friending system but is often ignored.
Secure communication channels are usually set up by
 authenticated key agreement protocols.
This can be performed by relying on a public-key infrastructure, e.g., based on
RSA or the Diffie-Hellman protocol \cite{diffie1976new}.
The public-key based methods allow parties to share authenticated information about each other,
 however they need a trusted third party.
Although Diffie-Hellman key exchange method allows two parties that
 have no prior knowledge of each other to jointly establish a shared secret key,
 it needs multiple interactions between two parties and is known to be vulnerable
 to the Man-in-the-Middle attack.

Device pairing is a another technique to generate a
 common secret between two devices that shared no prior secrets
 with minimum or without additional hardware.
Examples include the "resurrecting duckling" \cite{stajano2000resurrecting}, "talking-to-strangers" \cite{balfanz2002talking}, "seeing-is-believing" \cite{mccune2005seeing} and Short Authenticated Strings (SAS) based key agreement \cite{vaudenay2005secure} and \cite{pasini2006sas}.
However, they employ some out-of-band secure channel to exchange
 authenticated information or leverage the ability of users to authenticate each other by visual and verbal contact.
In addition, the interaction cost is still not well suited to decentralized mobile
 social networks where secure connections are needed between all parties at once.
With these existing schemes, it is more complicated to establish a group key.

\subsection{Attribute Based Encryption}

Attribute based encryption is designed for access control of
 shared encrypted data stored in a server.
Only the user possessing a certain set of credentials or attributes is able to access data.
It was introduced by Sahai and Waters \cite{sahai2005fuzzy}, and then later improved in \cite{goyal2006attribute,bethencourt2007ciphertext,chase2007multi}.
All the ABE schemes rely on asymmetric-key cryptosystem, which cost expensive computation.


%

\section{Discussion and Conclusion}
\label{sec:conclusion}

In this paper, we design a novel symmetric-encryption based privacy-preserving profile matching
 and secure communication channel establishment mechanism in decentralized social networks without
 any presetting or trusted third party.
We take advantage of the common attributes between matching users
 to encrypt a secret message with a channel key in it.
Several protocols were proposed for achieving different levels of privacy.
We rigorously compared the performances of our protocols with existing protocols and conducted extensive evaluations on the performances using a large scale dataset from real social networking.

{\small

}

\end{document}